\documentclass[twoside,a4paper,11pt]{sea10}
\usepackage{graphicx}
\usepackage{hyperref}
\usepackage{movie15}
\def\micron{{$\mu$m}}
\def\herschel{\textit{Herschel}}
\def\spitzer{\textit{Spitzer}}
\begin{document}
\pagenumbering{arabic}
\pagestyle{myheadings}
\thispagestyle{empty}
{\flushleft\includegraphics[width=\textwidth,bb=58 650 590 680]{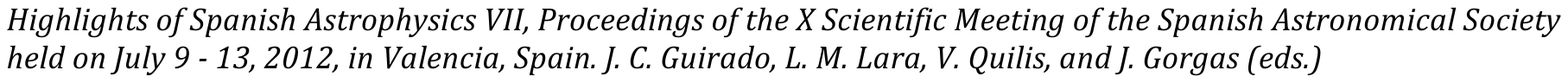}}
\vspace*{0.2cm}
\begin{flushleft}
{\bf {\LARGE
%
\vspace{-2cm}
SAFIR: testing the coexistence of AGN and star formation activity and the nature of the dusty torus in the local universe.\footnote{{\it Herschel} is an ESA space observatory with science instruments provided by European-led Principal Investigator consortia and with important participation from NASA.} 
%
}\\
\vspace*{1cm}
%
M. S\'anchez-Portal$^{1,2}$,
M. Castillo-Fraile$^{1,2}$,
C. Ramos Almeida$^{3}$,
P. Esquej$^{4,5}$,
A. Alonso-Herrero$^{5}$,
A. M. P\'erez Garc\'{\i}a$^{3}$,
J. Acosta-Pulido$^{3}$,
B. Altieri$^{1}$,
A. Bongiovanni $^{3}$,
J. M. Castro-Cer\'on$^{1}$, 
J. Cepa$^{3}$,
D. Coia$^{1}$,
L. Conversi$^{1}$,
J. Fritz$^{6}$,
J. I. Gonz\'alez-Serrano$^{5}$,
E. Hatziminaoglou$^{7}$,
M. Povi\'c$^{8}$,
J. M. Rodr\'{\i}guez Espinosa$^{3}$
and
I. Valtchanov$^{1}$
%
}\\
\vspace*{0.5cm}
%
$^{1}$
European Space Astronomy Centre (ESAC)/ESA, Madrid, Spain\\
$^{2}$
Ingenier\'{\i}a y Servicios Aeroespaciales, Madrid, Spain\\
$^{3}$
Instituto de Astrof\'{\i}sica de Canarias, La Laguna, Tenerife, Spain\\
$^{4}$
Centro de Astrobiolog\'{\i}a, INTA-CSIC, Madrid, Spain\\ 
$^{5}$
Instituto de F\'{\i}sica de Cantabria, CSIC-UC, Santander, Spain\\
$^{6}$
Sterrenkundig Observatorium, Universeit Gent,  Belgium\\
$^{7}$
European Southern Observatory, Garching bei M\"unchen, Germany\\
$^{8}$
Instituto de Astrof\'{\i}sica de Andaluc\'{\i}a, Granada, Spain\\
%
\end{flushleft}
%
\markboth{
SAFIR: testing AGN and star formation activity.
}{ 
%
S\'anchez-Portal et al.
%
}
\thispagestyle{empty}
\vspace*{0.4cm}
\begin{minipage}[l]{0.09\textwidth}
\ 
\end{minipage}
\begin{minipage}[r]{0.9\textwidth}
\vspace{1cm}
\section*{Abstract}{\small
%
We present the Seyfert and star formation Activity in the Far-InfraRed (SAFIR) project, a small (15.1h) \herschel\ guaranteed time proposal performing PACS and SPIRE imaging of a small sample of nearby Seyfert galaxies.
This project is aimed at studying the physical nature of the nuclear IR emission by means of multi-component spectral energy distribution (SED) fitting and the star formation properties of AGN hosts, as traced by cold dust. We summarize the results achieved so far and outline the on-going work. 
%
\normalsize}
\end{minipage}
%
%
%
\section{Introduction \label{intro}}

The infrared (IR) range provides a unique probe into the coeval AGN and starburst phenomena. Most of the mid-IR (MIR) and far-IR (FIR) emission comes from dust, being of thermal origin. According to the unified model, the central engine and broad-line region (BLR) are obscured by a thick dust torus. The grains absorb the UV/optical photons and re-radiate in the IR. The dusty torus emission peaks in the MIR (7-30\,\micron), being well characterized by existing facilities (eg. \spitzer\ , T-ReCS). It extends to the FIR, where the star formation (SF) related emission becomes dominant. So far this has been poorly constrained, due to the limited spatial resolution and spectral coverage.
According to \cite{perez01},  the MIR and FIR SED of Seyfert galaxies can be explained solely through
thermal re-radiation of higher energy photons by dust; this thermal emission is made up of three different components:
\textit{(a)} warm dust (120-170 K), heated by the AGN; \textit{(b)} cold dust (40-70 K), heated by star formation, and \textit{(c)} very cold dust (15-25 K), typical of dust heated by the general interstellar radiation field.
\herschel\ \cite{herschel} offers a unique window to study the FIR/sub-mm emission from nearby galaxies: on the one hand, the PACS \cite{pacs} photometer permits imaging at 70, 100 and 160\,\micron\ bands with unprecedented spatial resolution ($\sim$\,5.5\,arcsec at 70\,\micron), thus allowing to characterize the SED minimizing the contamination from the host galaxy;  and on the other,  the SPIRE photometer allows to sample the formerly unexplored region in three bands at 250, 350 and 500\,\micron\  with a relatively high spatial resolution, probing the cold and very cold dust components across the galaxy and even in the nuclear and circum-nuclear region.
These data can be used to build SEDs sampling the peak and Rayleigh-Jeans tail of the thermal cold/very cold dust emission, allowing to derive masses and temperatures. The star formation rate (SFR) can be also derived.
The spatial sampling also allows distinguishing different regions within the galaxies (e.g. nucleus, arm, inter-arm regions). The two main subjects of study within SAFIR are: \textit{(a)} The dusty torus: the current models fall in two main categories: smooth and clumpy distributions.  When applying any AGN torus model to the actual SEDs of Seyfert galaxies, three ingredients should be considered: AGN, starburst and host galaxy. The starburst contribution is mainly determined by the FIR data points. Moreover, the torus and starburst emissions overlap smoothly in the FIR. Therefore, the availability of FIR data with high spatial resolution and wide spectral coverage is of crucial importance to break the model degeneracy. \textit{(b)} Study of star formation and nuclear activity: the interrelationship between star formation and accretion onto massive black holes is crucial to understanding galaxy formation and evolution. \herschel\  provides a unique means to map the star formation activity across the host galaxies of AGN, allowing to to determine the location and extent of the (obscured) star forming regions and the variation of the SFR with the radial distance, and permitting an accurate mapping of the the dust composition and temperature.

\section{Sample of galaxies and technical implementation \label{implementation}}

The SAFIR sample 
comprises 18 nearby galaxies that were selected in such a way that different nuclear classes (Seyfert 1.x \& Seyfert 2)  were represented. Ten objects are barred spirals/lenticular galaxies. Five objects are peculiar/interacting systems. Four galaxies are confirmed Luminous or Ultra-luminous IR galaxies (LIRG/ULIRG). Only objects with either high-resolution ground-based MIR or \textit{Spitzer} data were chosen.  In addition, all the targets have optical, NIR, X-ray and radio data. Therefore, a complete multi-wavelength SED can be constructed for all targets.

The observations were carried out using  PACS and SPIRE scan map modes. The mapped areas for all targets were adjusted in such a way that the host galaxy and a background region fit within the surveyed area.  
%
 The 1$\sigma$ sensitivities achieved by the PACS photometer were approximately 3.6, 3.9 and 3.9 mJy/beam at 70, 100 and  160\,$\mu$m, respectively.  For the SPIRE photometer,  1$\sigma$ sensitivities just above the extragalactic confusion limit were achieved: 5.5, 7.6 and 6.4 mJy/beam at 250, 350 and 500 $\mu$m. 
 These sensitivities allowed mapping, not only the nuclear and circum-nuclear regions but also large areas within the spiral discs.

\section{Results \label{results}}

In this section we present some already published results from the SAFIR project \cite{ramos11,esquej12,alonso12} and on-going work.

\subsection{NGC 3081\label{n3081}}

\begin{figure}
\center
\hspace{-1cm}
\includegraphics[scale=0.9]{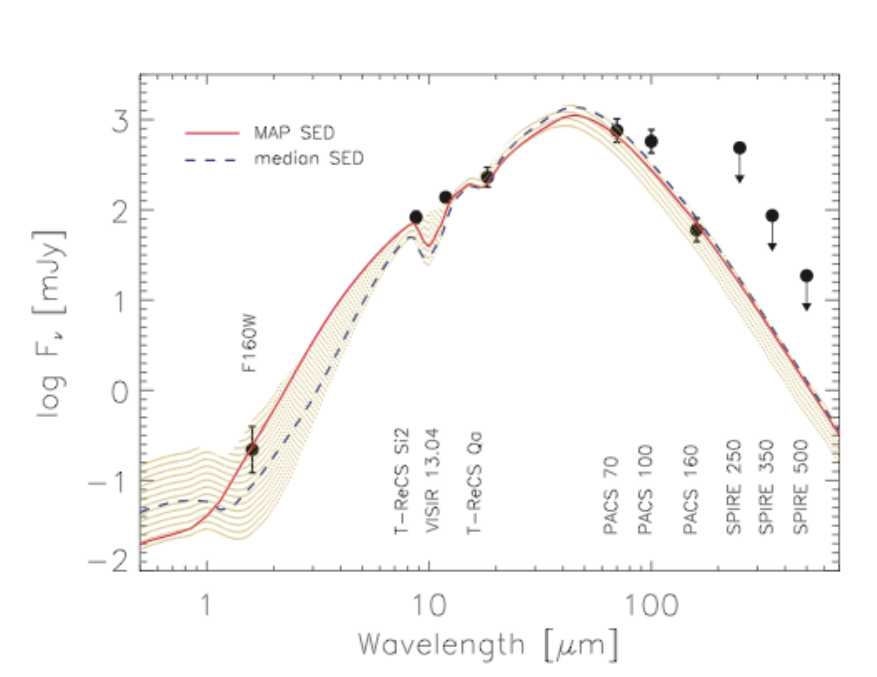}  ~
\includegraphics[scale=0.86]{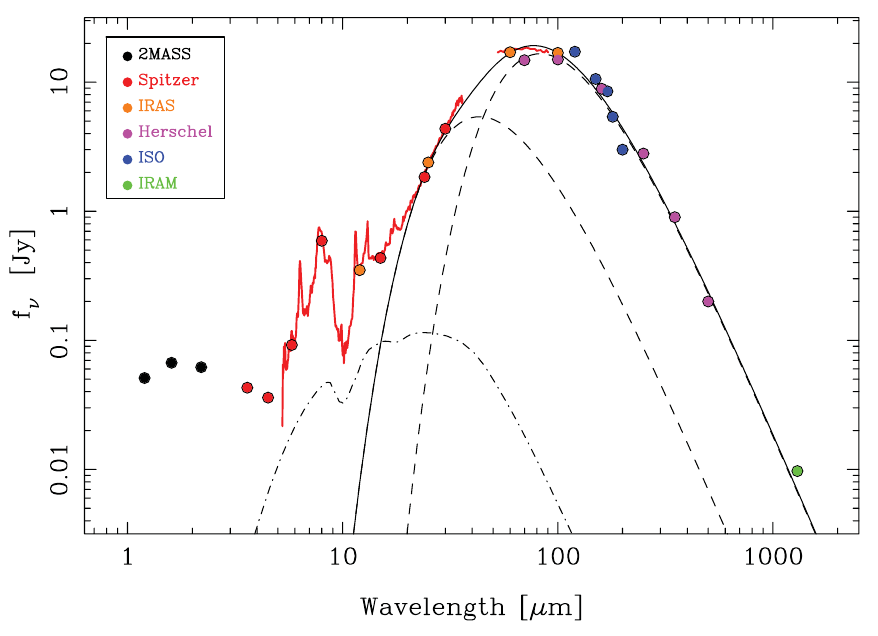} 

\caption{\label{fits}\footnotesize Left: IR SED of NGC~3081 from \cite{ramos11}. Solid and dashed lines are the Ôbest fitÕ to the data (MAP) and the model described by the median of the posteriors, respectively. The shaded region indicates the range of models compatible with the observations at the 2$\sigma$ level. The SPIRE nuclear fluxes include emission from the r1 ring, and therefore have been considered as upper limits in the fit. Right: IR SED of Mrk~938 from \cite{esquej12}, including modeling to the MIR and FIR data using two modified blackbodies (solid black line), with  T$_{w}$\,=\,67\,K and T$_{c}$\,=\,35\,K (dashed lines). The grey dot-dashed line is the fitted type 2 AGN template. 
}
\end{figure}

NGC~3081 is  an early-type barred spiral galaxy ((R)SAB0/a(r)), harboring a Seyfert 2 nucleus. A series of star-forming nested ring-like features are observed:
nuclear (r1, 2.3 kpc), inner (r2, 11 kpc), outer (26.9 kpc). Our study combines  \textit{Herschel} FIR data with high-resolution ground-based NIR/MIR data \cite{ramos11}. The r2 ring is clearly resolved in our images up to 250\,\micron. The nuclear SED was computed with unresolved FIR fluxes (r\,$\leq$\,1.7\,kpc) plus  NIR and MIR data as shown in Fig. \ref{fits} (left panel). The torus was modeled using a clumpy model \cite{nenkova08}. We wanted to test how the inclusion of FIR data affects the torus parameters, specially the torus size. We found that the inclusion of FIR data results in a notable increase of the torus outer radius: R$_o$\,=\,4$^{+2}_{-1}$\,pc vs. R$_o$\,=\,0.7\,$\pm$\,0.3\,pc obtained using only NIR and MIR data \cite{ramos11a}. The radial distribution of clouds (as given by the clumpy model q parameter) also flattens when FIR data are included in the modeling (q\,=\,0.2 vs q\,=\,2.3). The remaining fitting parameters (width of the angular distribution, number of clouds, inclination angle and cloud optical depth) are in agreement with those obtained without the inclusion of the FIR range.  At larger scales (1.7\,kpc\,$\leq$\,r\,$\leq$\,5.4\,kpc), the FIR emission can be reproduced by cold dust at T\,=\,28\,$\pm$\,1\,K (assuming a modified blackbody with emissivity $\beta$\,=\,2), heated by young stars within the galaxy disc (likely located at r1). Finally, the FIR emission from the outer part of the galaxy is compatible with dust heated by the interstellar radiation field (T\,=\,19\,$\pm$\,3\,K ).

\subsection{Mrk 938\label{mrk938}}

The morphologically peculiar galaxy Mrk~938  has been proposed to be the remnant of a gas-rich merger of two unequal mass galaxies \cite{schweizer}.  According to its IR luminosity is a LIRG. The galaxy hosts a Seyfert 2 AGN and is known for a vigorous starburst activity. We have performed a  multiwavelength study, including X-ray, NIR, MIR and \herschel\ FIR data \cite{esquej12}.
The decomposition of the MIR \textit{Spitzer}/IRS spectrum shows that the AGN bolometric contribution to the MIR and total IR luminosity is small [L$_{bol}$(AGN)/L$_{IR}$\,$\sim$\,0.02], which agrees with previous estimations. We have characterized the physical nature of its strong IR emission. The MIPS 24\,\micron\ and PACS 70\,\micron\ images reveal that the bulk of the star formation activity is located in a compact, obscured region of $\leq$\,2 kpc. We have used \herschel\ imaging data for the first time to constrain the cold dust emission with unprecedented accuracy. We have fitted the integrated IR SED (see Fig. \ref{fits}, right panel) and derived the properties of the dust. We have found that the MIR to FIR spectrum can be properly modeled with two modified blackbodies, with fixed emissivity $\beta$\,=\,2 and best-fitting temperatures of T$_{w}$\,=\,67\,K and T$_{c}$\,=\,35\,K for the warm and cold dust dust
components, respectively. The FIR-only spectrum can be fitted to a single modified blackbody with  $\beta$\,=\,2 and T\,=\,36.5\,K. Using this value and the SPIRE 250\,\micron\ flux, we have derived the dust mass using the expression from \cite{hildebrand} assuming an absorption coefficient  $\kappa_{250\mu m}$\,=\,4.99\,cm$^2$g$^{-1}$. The obtained value, M$_{dust}$\,=\,3\,$\times$\,10$^7$\,M$_{\odot}$, is consistent with values derived for local ULIRGs and other IR-bright galaxies.

\subsection{NGC 1365\label{n1365}}

NGC~1365 (Fig. \ref{temp_maps}, top right) is a supergiant barred spiral galaxy (SB(s)b). It is a nearby (18.6 Mpc) LIRG harboring a Seyfert 1.5 type nucleus. The inner Linblad resonance (ILR) region of the galaxy  contains a powerful nuclear starburst ring with an approximate diameter of 2\,kpc. We have probed the nuclear and circumnuclear activity of this galaxy in the IR \cite{alonso12}. The strong star formation activity in the ring is resolved by the \herschel/PACS imaging data that shows some substructures (super star clusters), as well as by the \spitzer\  24\,\micron\  continuum emission, [Ne {\sc ii}]12.81\,\micron\  line emission, and 6.2 and 11.3\,\micron\ PAH emission. The active galactic nucleus (AGN) is the brightest source in the central regions up to $\lambda$\,$\sim$\,24\,\micron, but it becomes increasingly fainter in the FIR when compared to the emission originating in the IR clusters located in the ring. We modeled the AGN unresolved IR emission with the clumpy torus models and estimated that the AGN contributes only to a small fraction ($\sim$\,5\%) of the IR emission produced in the inner $\sim$\,5 kpc. The torus size, estimated using the clumpy torus model \cite{nenkova08} is $\sim$\,5\,pc. We fitted the non-AGN 24--500\,\micron\ spectral energy distribution of the ILR region and found that the dust temperatures and mass are similar to those of other nuclear and circumnuclear starburst regions. Finally, we show that the comparison of the IR SFR with that derived from H$\alpha$ indicates that $\sim$\,85\% of the on-going star formation within the ILR is taking place in dust--obscured regions.

\section{On-going work\label{ongoing}}

\begin{figure}
\center
\includegraphics[scale=0.5]{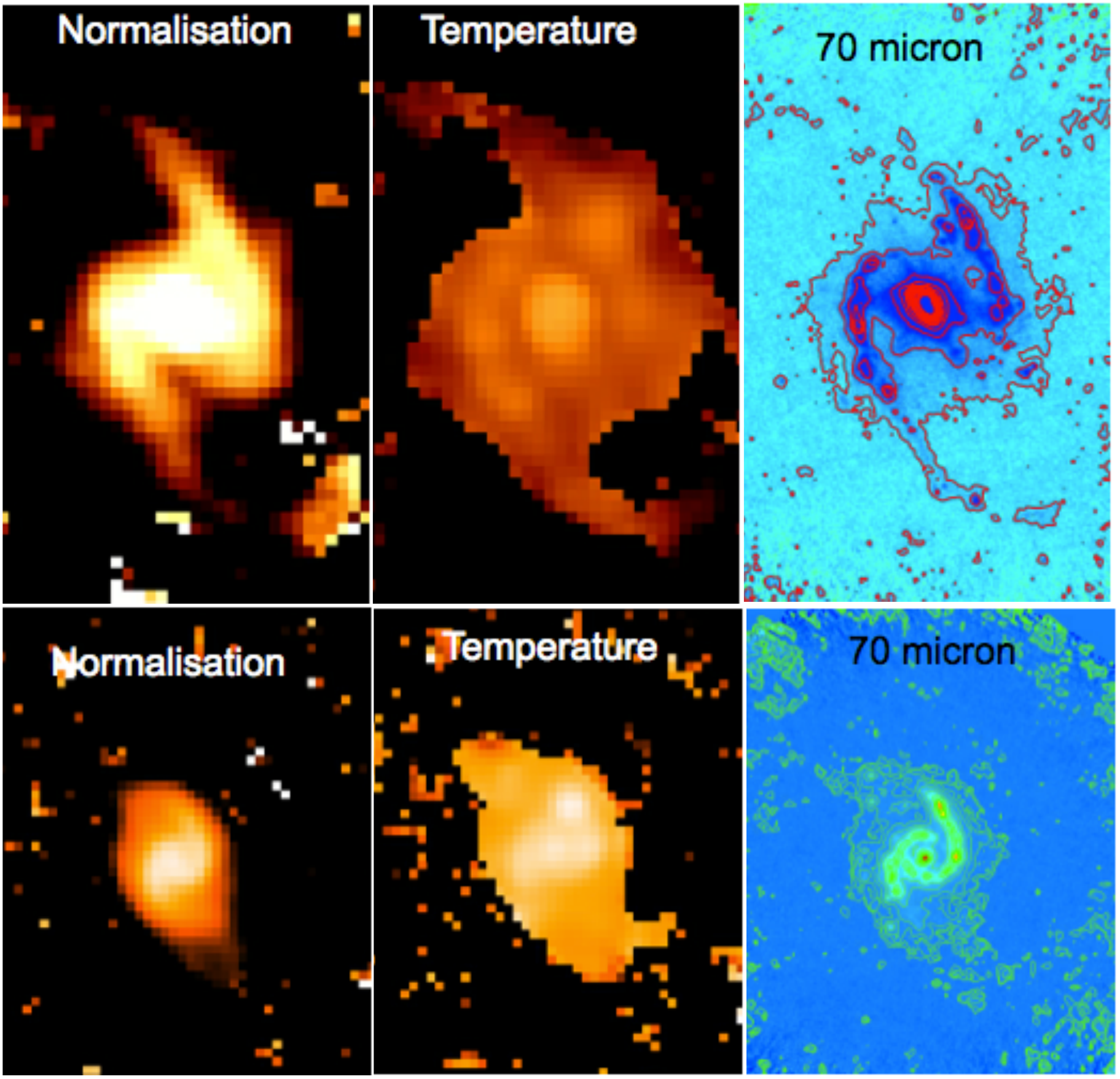}  

\caption{\label{temp_maps}\footnotesize  Temperature maps of NGC~1365 (top) and NGC~1566 (bottom). 
Average temperatures range from $\sim$\,17--18\,K in the inter-arm regions to T\,$\sim$23--24\,K  in the bright spots within the spiral arms. The highest average dust temperatures are observed in the central region of NGC~1365 with T\,$\sim$26\,K.}
\end{figure}

We have started a study of the dust properties of spatially well-resolved AGN hosts (S\'anchez-Portal, Castillo-Fraile et al. in preparation). Within our sample, four galaxies (NGC~1365, NGC~4258, NGC~1566 and NGC~5728) have an apparent size large enough to allow a detailed analysis of the spatial dust properties, notably its temperature and mass that can be directly compared with the star formation characteristics. For these objects, we are exploiting the spatial resolution of our observations to produce dust temperature maps. 
Assuming an optically thin emission, the flux density can be expressed as $f_{\nu} \propto \nu^{\beta} B(\nu, T_{dust})$ where $\beta$ is the dust emissivity. As already stated, several dust components with different temperature should be generally considered, so the flux density can be expressed as:

\begin{equation}
f_{\nu}(\lambda) = \sum_{i=1}^{n}\frac{N_i}{\lambda^{\beta + 3} (e^{hc/\lambda k T_i} - 1)}
\end{equation}

\noindent where N$_i$ are the normalization constants and T$_i$ are the temperatures of the different components. 
The procedure devised to generate temperature maps includes the following steps: after a standard reduction procedure (see \cite{alonso12} for a description), the PACS 70, 100 and 160\,\micron\ and SPIRE 250 and 350\,\micron\ maps have been convolved to the resolution of the SPIRE 500\,\micron\ images and resampled to the largest pixel size (that of the SPIRE\,500\,\micron\ maps, set to 14\,arcsec). The images have been spatially registered and used as input to an IDL procedure that performs a least-squares fit to either one or two dust components (n\,=\,1, 2) at each pixel. In the maps shown in Fig. \ref{temp_maps} we have used a single temperature component with a  fixed emissivity $\beta$\,=\,2 to create the temperature maps of NGC~1365 (top) and NGC~1566 (bottom). The latter is a bright (LC II-III), nearby (11.83 Mpc) SAB(s)bc spiral galaxy harboring a Seyfert type 1.5 nucleus. By comparison with the morphology of 70\,\micron\ and optical H$\alpha$ images, it can be observed how closely the temperature maps follow the star formation regions, with the highest temperatures corresponding to areas of high SF activity. 

\section{Conclusions}

The high spatial resolution \herschel\ PACS \& SPIRE observations are demonstrating the importance of the FIR to improve the quality and reliability of the AGN torus fits. Parameters as important as the torus radius and cloud radial distribution have a strong dependency of this spectral range. Moreover, the FIR data are crucial to characterise the starburst contribution and to constrain the dust properties. The quality of the \herschel\ data is allowing us to study the spatial distribution of dust within the galaxies, thus permitting  to characterize the variation of dust properties (e.g. temperature) and SFR with the nuclear distance.

\small  
%
\section*{Acknowledgments}   
%

\footnotesize
We would like to acknowledge the \herschel\  Project Scientist, G\"oran Pilbratt, for making possible the implementation of this project kindly providing the required guaranteed time from the PS budget.



%
\end{document}